\documentclass[a4paper]{article}

\usepackage{INTERSPEECH2021}
\usepackage{multirow}
\usepackage{makecell}
\usepackage{footmisc}
\usepackage{amsfonts}
\usepackage{arydshln}

\title{Text Anchor Based Metric Learning for Small-footprint Keyword Spotting}
\name{Li Wang$^1$, Rongzhi Gu$^1$, Nuo Chen$^1$, Yuexian Zou$^{1,2,*}$\thanks{*Corresponding author}}
\address{
  $^1$ ADSPLAB, School of ECE, Peking University, Shenzhen, China\\
  $^2$ Peng Cheng Laboratory, Shenzhen, China}
\email{\{1901213145, 1701111335, nuochen, zouyx\}@pku.edu.cn}

\begin{document}

\maketitle
\begin{abstract}
Keyword Spotting (KWS) remains challenging to achieve the trade-off between small footprint and high accuracy. Recently proposed metric learning approaches improved the generalizability of models for the KWS task, and 1D-CNN based KWS models have achieved the state-of-the-arts (SOTA) in terms of model size. However, for metric learning, due to data limitations, the speech anchor is highly susceptible to the acoustic environment and speakers. Also, we note that the 1D-CNN models have limited capability to capture long-term temporal acoustic features. To address the above problems, we propose to utilize text anchors to improve the stability of anchors. Furthermore, a new type of model (LG-Net) is exquisitely designed to promote long-short term acoustic feature modeling based on 1D-CNN and self-attention. Experiments are conducted on Google Speech Commands Dataset version 1 (GSCDv1) and 2 (GSCDv2). The results demonstrate that the proposed text anchor based metric learning method shows consistent improvements over speech anchor on representative CNN-based models. Moreover, our LG-Net model achieves SOTA accuracy of 97.67\% and 96.79\% on two datasets, respectively. It is encouraged to see that our lighter LG-Net with only 74k parameters obtains 96.82\% KWS accuracy on the GSCDv1 and 95.77\% KWS accuracy on the GSCDv2.
\end{abstract}

\section{Introduction}
\label{sec:intro}
With speech technology development, speech assistants can help people solve affairs more efficiently, such as querying weather and controlling air conditioning.
People increasingly enjoy the convenience of the hands-free experience. Keyword Spotting (KWS) is the beginning of the human-computer speech interaction, aims at distinguishing between each of the target keywords of interest and non-target sounds such as general speech (non-target words) and noises \cite{huh2020metric}.

Recently, researchers have improved the generalization performance of models in terms of data enhancement \cite{raju2018data}, loss function \cite{liu2019loss, zhang2020re}, automatic gain control \cite{prabhavalkar2015automatic}, negative sample mining~\cite{hou2020mining}, and metric learning \cite{huh2020metric,sacchi2019open}. In particular, the metric learning based on triplet loss \cite{huh2020metric,sacchi2019open} has demonstrated excellent performance.
The KWS models are trained using triplet consists of an $anchor$ sample, a $positive$ sample from the same class with the $anchor$, and a $negative$ sample from a different class. The objective of the network training is to minimize the distance between the embeddings of the $anchor$ and the $positive$ sample while maximizing the distance between the embeddings of the $anchor$ and the $negative$ sample.
It can be seen that the performance is much dependent on the choice of the anchor. However, due to data limitations, the speech anchor is highly susceptible to the acoustic environment and speakers, making the anchor embedding more variant during training. In addition, deviated speech anchors may make the KWS models suffering from a local optimality issue, which leads to the degradation of the performance. 

\begin{figure*}
\centerline{\includegraphics[width=2\columnwidth]{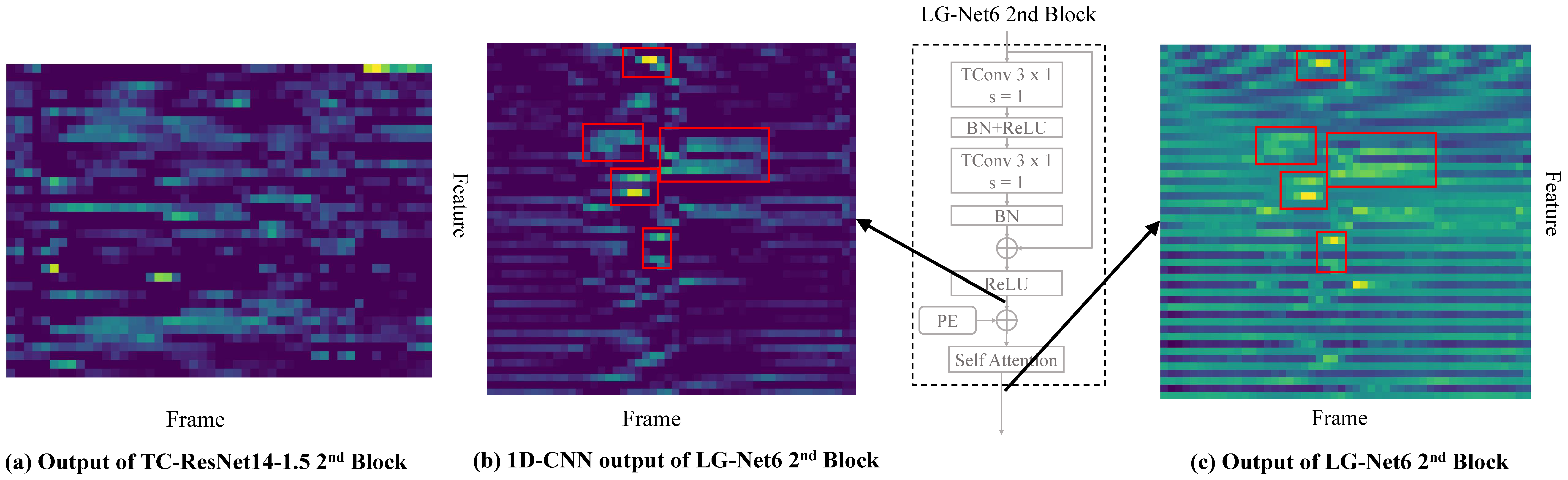}}
\caption{Feature maps produced by the TC-ResNet14-1.5 and proposed model LG-Net using a benchmark image, saliency for each feature is represented by brightness. (a) and (b) shows that the feature maps outputs by 1D-CNN are sparse, representing that 1D-CNN focuses on short-term temporal information. (c) shows the self-attention layer modeling the long-term information while preserving the short-term temporal information captured by the 1D-CNN.}
\label{visual}
\end{figure*}

On the other hand, Deep neural networks (DNNs) have recently proven to yield efﬁcient small-footprint solutions for KWS \cite{chen2014small, sainath2015convolutional, tang2018deep, shan2018attention, zhang2018sequence, choi2019temporal, bai2019time, li2020small, mo2020neural}.
In particular, more advanced architectures, such as Convolutional Neural Networks (CNNs), have been applied to solve KWS problems under limited memory footprint as well as computational resource scenarios, showing excellent accuracy.
Most CNN-based KWS models receive features, such as Mel-Frequency Cepstral Coefﬁcient (MFCC), as a 2D input. However, such 2D CNN-based KWS models struggle with capturing the dependency between low and high frequencies with the relatively shallow network. To address this problem, \cite{choi2019temporal, li2020small} utilize temporal convolution to capture low and high-frequency features in a shallow network, which achieves the best KWS performance.
Despite their success, due to local region perception and weight sharing characteristics of CNNs, the long-term temporal information may not be considered. Figure \ref{visual}a shows the feature maps produced by a representative 1D-CNN-based KWS model, TC-ResNet14-1.5~\cite{choi2019temporal}. It is noted that the saliency regions are distributed sparsely, which implies that 1D-CNN focuses on the short-term temporal information. In fact, it is crucial to capture the long-term temporal information for KWS, considering that the characteristics of keywords are usually different on the time scale.

In this study, we aim at boosting the performance of the small-footprint KWS model. To improve the stability of the anchor, we replace the speech anchor with the text anchor and propose a text anchor based metric learning method for KWS. Speciﬁcally, target keywords and non-target words are encoded in the text anchors by pretraining models (like BERT~\cite{devlin2018bert}).
Moreover, the local-global network (LG-Net) is designed to model local and global information.
Speciﬁcally, the self-attention layer \cite{vaswani2017attention} is stacked to the 1D temporal convolution layer to capture the long-term and short-term temporal information.
Experimental results demonstrate that the proposed text anchor based metric learning method shows consistent improvements over speech anchor on representative CNN-based models. Besides, our LG-Net model achieves SOTA accuracy of 97.67\% and 96.79\% on Google Speech Commands Dataset version 1 (GSCDv1) and version 2 (GSCDv2), respectively.

\section{Proposed Method}
\label{sec:pro}

\subsection{Data Process}
The raw audio is decomposed into a sequence of frames where the window length is 25 ms and the stride is 10 ms for feature extraction. We use 40 Mel-Frequency Cepstral Coefficient (MFCC) features for each frame and stack them over the time-axis, denote as $M \in \mathbb{R}^{T \times F}$ where $F$ represents the dimension of the MFCC feature, and $T$ denotes the number of frames. The text anchor is extracted by BERT-base \cite{devlin2018bert} denoted as $V \in \mathbb{R}^{D}$, where $D$ represents the dimension of the text anchor (\textit{i.e.} $D=768$).

\subsection{Text anchor based metric learning method}
\label{sec:pro:text}
\begin{figure}
\centerline{\includegraphics[width=\columnwidth]{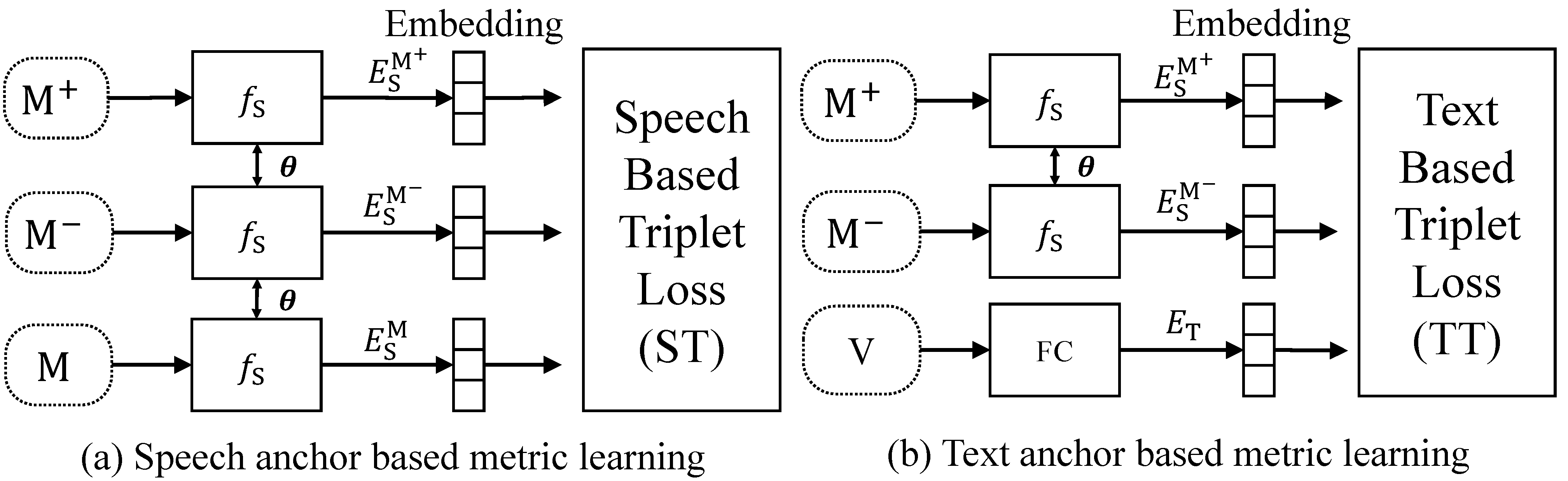}}
\caption{(a) Speech anchor based metric learning and (b) proposed text anchor based metric learning. $M^{\mathrm{+}}$ and $M^{\mathrm{-}}$ are input MFCCs from different classes, $M$ is input MFCC from the same class of $M^{\mathrm{+}}$, and $V$ as the input text vector encoded by BERT \cite{devlin2018bert} corresponding $M^{\mathrm{+}}$.
$E_{\mathrm{S}}$ represents the speech embedding output from the speech embedding extraction module $f_\mathrm{S}$. $E_{\mathrm{T}}$ represents the text embedding.}
\label{framework}
\end{figure}

As mentioned in Section \ref{sec:intro}, the susceptibility of the speech anchor leads to the poor performance of the model. Therefore, we propose a text anchor based metric learning method for KWS, illustrated in Figure \ref{framework}b.
The input triplet is denoted as ($M^{\mathrm{+}}$, $M^{\mathrm{-}}$, $V$), where $M^{\mathrm{+}}$ and $M^{\mathrm{-}}$ are input MFCCs from different classes, and $V$ as the input text vector corresponding to $M^{\mathrm{+}}$. 
The speech embeddings $E_{\mathrm{S}}^{M^{\mathrm{+}}} \in \mathbb{R}^{D^{\prime}}$ and $E_{\mathrm{S}}^{M^{\mathrm{-}}} \in \mathbb{R}^{D^{\prime}}$ are extracted from $M^{\mathrm{+}}$ and $M^{\mathrm{-}}$ by the speech embedding extraction module $f_\mathrm{S}$, respectively, where $D^{\prime}$ represents the dimension of the speech embedding (\textit{i.e.} $D^{\prime}=128$).
In parallel, $V$ is mapped as a text embedding $E_{\mathrm{T}} \in \mathbb{R}^{D^{\prime}}$. The triplet loss is employed to decreases the distance between the  $E_{\mathrm{T}}$ and $E_{\mathrm{S}}^{M^{\mathrm{+}}}$  and increases the distance between $E_{\mathrm{T}}$ and $E_{\mathrm{S}}^{M^{\mathrm{-}}}$.
For a single sample, the triplet loss is thus
\begin{equation}
L_{t r i}=\max \left(d\left(E_{\mathrm{S}}^{M^{\mathrm{+}}}, E_{\mathrm{T}}\right)-d\left(E_{\mathrm{S}}^{M^{\mathrm{-}}}, E_{\mathrm{T}}\right)+\alpha, 0\right)
\end{equation}
where $d(x, y)=\|x-y\|_{p}$ is the pairwise-distance between x and y (\textit{i.e. $p=2$}); $\alpha$ is a constant margin (\textit{i.e.} $\alpha$ = 1).

Note that the phone embeddings extracted by the text encoder in \cite{sacchi2019open} is fundamentally different from the text anchor proposed in this paper.
In \cite{sacchi2019open}, a pronunciation dictionary is used to map words to their sequence of phonemes.
Therefore, the phone embeddings do not contain the information about other words, although it is not influenced by factors such as acoustic environment and speaker.
The text anchor in this paper is extracted from the BERT \cite{devlin2018bert} which is trained on the large-scale text corpus, it contains information about other words.

\subsection{Local-global Neural Network}
\label{LGNet}
\begin{figure}
\centerline{\includegraphics[width=\columnwidth]{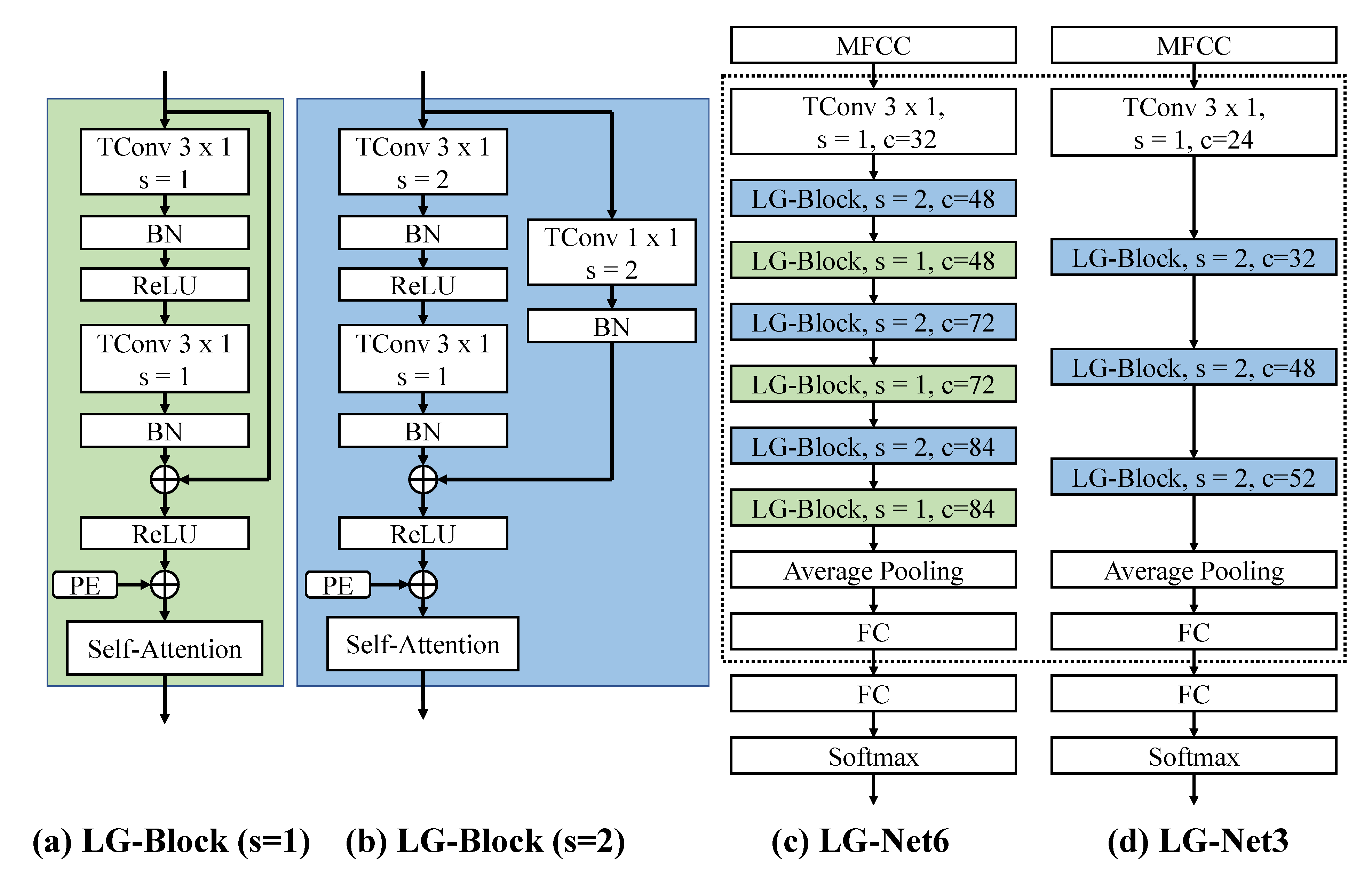}}
\caption{The local-global block (LG-Block) when (a) stride = 1 and (b) stride = 2 and two implementations of local-global neural network (LG-Net) when (c) LG-Net6 and (d) LG-Net3.
TConv means the temporal convolution layer, ``s'' and ``c'' indicates stride and channel of TConv, respectively.
PE means the position encoding \cite{vaswani2017attention}.
BN and FC denote batch normalization and fully connected layer. The part inside the dotted line is regarded as the speech embedding extraction module.
}
\label{Net}
\end{figure}

As mentioned in Section \ref{sec:intro}, for KWS, long-term temporal information is crucial because the characteristics of keywords are usually different on the time scale. In this work, we utilize the self-attention \cite{vaswani2017attention} layer to capture long-term temporal information.
The temporal convolution with residual structure \cite{choi2019temporal} is adopt in this paper, which is one of the most widely used 1D-CNN architectures for KWS.
To reduce the number of parameters, $3 \times 1$ kernels are utilized instead of $9 \times 1$ kernels. As Figure \ref{Net} shows, and the self-attention layer is stacked to the 1D-CNN layers, denoted as LG-Block. Following the design of the previous study \cite{choi2019temporal}, the identity shortcuts can be directly used when the input and output have matching dimensions (Figure \ref{Net}a); otherwise, we use an extra $1 \times 1$ convolution with a stride to match the dimensions (Figure \ref{Net}b). Local-global neural network (LG-Net) can be easily constructed by stacking LG-Blocks. We select LG-Net6 (Figure \ref{Net}c), which has 6 LG-Blocks as our base model. To establish a small-footprint model, we compress the LG-Net6 by taking out three LG-Blocks and reducing the number of channels, named LG-Net3.

Considering the variation in feature maps due to speaking style, speed of speech, we use an averaging pooling strategy on the time dimension. The output of the average pooling layer is given to a fully connected layer to generate speech embedding $E_{\mathrm{S}} \in \mathbb{R}^{D^{\prime}}$.

\begin{equation}
E_{\mathrm{S}}=f_{\mathrm{S}}\left(M ; \theta_{\mathrm{S}}\right)
\end{equation}
where $\theta_{\mathrm{S}}$ denote the parameters of the $f_\mathrm{S}$. 

Finally, the predicted score, $\hat{y} \in \mathbb{R}^{C}$ where $C$ denote the number of class, is obtained from $E_{\mathrm{S}}$
\begin{equation}
\hat{y}=\sigma\left(f_{\mathrm{FC}}\left(E_{\mathrm{S}} ; \theta_{\mathrm{FC}}\right)\right)
\end{equation}
where $\sigma$ is the sigmoid function; $\theta_{\mathrm{FC}}$ denotes the parameters of the fully connected layer $f_{\mathrm{FC}}$.
KWS as a multi-classification problem, the cross entropy (CE) loss is adopted to optimize the model, for the given ground truth $y \in \mathbb{R}^{C}$ (where $y^{(i)}=\{0,1\}$ denotes whether label $i$ appears or not), the loss $L$ is calculated using binary cross-entropy:
\begin{equation}
L_{CE}=\sum_{i=1}^{C} y^{(i)} \log \left(\hat{y}^{(i)}\right)+\left(1-y^{(i)}\right) \log \left(1-\hat{y}^{(i)}\right)
\end{equation}

In this work, the networks are optimized by the loss $L$, which is calculated by weighted sum of $L_{\mathrm{tri}}$ and $L_{\mathrm{CE}}$
\begin{equation}
L=\beta L_{\mathrm{tri}}+(1-\beta) L_{\mathrm{CE}}
\end{equation}
where $\beta$ is the balanced weight between $L_{\mathrm{tri}}$ and $L_{\mathrm{CE}}$.

Except for the cases of target keywords and non-target words, a ``silence'' class is considered to indicate the case when no speech is detected \cite{warden2018speech}. However, in this work, the text vector of ``silence'' does not match the definition in KWS, so the model cannot be trained directly using the text vector of ``silence''. To address the problem, we first train the network without the ``silence'' sample and then finetune the two FC layers on the full dataset with only CE loss.

\section{Experiments}
\label{sec:exp}
\subsection{Datasets}

The proposed models are trained and evaluated with Google Speech Commands Dataset version 1 (GSCDv1\footnote{\label{v1}http://download.tensorflow.org/data/speech\_commands\_v0.01.tar.gz}) and version 2  (GSCDv2\footnote{\label{v2}http://download.tensorflow.org/data/speech\_commands\_v0.02.tar.gz}) \cite{warden2018speech}. Following the implementation of Google~\cite{warden2018speech}, we seek to discriminate among 12 classes for GSCDv1: ``yes'', ``no'', ``up'', ``down'', ``left'', ``right'', ``on'', ``off'', ``stop'', ``go'', unknown, or silence, and 16 classes for GSCDv2 which add 4 keywords of GSCDv1: `backward", ``forward", ``follow", and ``learn".
In order to generate background noise, we randomly sample and crop background noises provided in the dataset. For a fair comparison, in our test set, the ``silence'' class test samples are taken from open source speech commands dataset test set version 2\footnote{http://download.tensorflow.org/data/speech\_commands\_test\_set\_v0.02.tar.gz} \cite{warden2018speech}, and test samples of other classes are written in the ofﬁcially released \textit{testing.list}\footref{v1}\footref{v2}. Table \ref{tab:dataset} shows the number of samples in training, validation, and test sets. Note that to make it easy for other researchers to reproduce the model and results in this paper, we do not use any data augmentation methods.

\begin{table}
  \centering
  \caption{The number of samples in training, validation and test set of google speech commands dataset version 1 and version~2.}
    \begin{tabular}{|p{30pt}|p{30pt}<{\centering}|p{30pt}<{\centering}|p{30pt}<{\centering}|p{30pt}<{\centering}|}
    \hline
    Dataset & Train & Valid & Test & Total\\
    \hline
    GSCDv1 & 55287 & 7218 & 7242 & 69747\\
    GSCDv2 & 89043 & 10401 & 11413 & 110857\\
    \hline
    \end{tabular}%
  \label{tab:dataset}%
\end{table}%

\subsection{Implementation Details}
Our implementation was done with Pytorch \cite{paszke2017automatic} deep learning toolkit.

\textbf{Training.}
All the models are trained with a mini-batch of 256 samples using stochastic gradient descent with weight decay of 0.001 and momentum of 0.9. For formula (5), the $L_{\mathrm{tri}}$ and $L_{\mathrm{CE}}$ weights to be the same, \textit{i.e.} $\beta= 0.5$, $L = 0.5L_{\mathrm{CE}}+0.5L_{\mathrm{tri}}$.
The initial learning rate is set to be 0.01 and decayed by a factor of 3 when the validation accuracy does not increase for 3 epochs. The training is terminated when validation accuracy does not increase for 10 epochs. For the fine-tuning stage mentioned in Section \ref{LGNet}, the training setup is the same as the text anchor based metric learning method, except that $\beta$ is set as 0.

\textbf{Evaluation.} The primary metric in our experiments is classiﬁcation accuracy, the proportion of correct decisions out of the total number of samples. We also report the number of parameters and the false reject rate (FRR) at 0.5\% false alarm rate (FAR). We train each model 10 times and report its average performance.

\subsection{Baselines}
To demonstrate the consistent performance improvement of our proposed text anchor based metric learning method, we reproduce CnnTradFpool3 \cite{sainath2015convolutional}, CnnOneFstride4 \cite{sainath2015convolutional}, TDNNSWSA~\cite{bai2019time}, TC-ResNet8 \cite{choi2019temporal}, and TC-ResNet14-1.5~\cite{choi2019temporal}, which are representative CNN-based KWS models.

We use three loss functions for training and comparison. First, the CE loss, which is the most widely used loss function in the KWS task. Second, the triplet loss based on speech anchor (\textit{ST}) is used to evaluate the performance gains from the triplet loss. As shown in Figure \ref{framework}a, the input triplet is denoted as ($M^{\mathrm{+}}$, $M^{\mathrm{-}}$, $M$), where $M$ is input MFCC randomly sampled from the training set with the same class of $M^{\mathrm{+}}$. Third, the triplet loss based on text anchor (\textit{TT}). As mentioned in Section \ref{LGNet}, we employ CE loss accompanied by the ST and the TT, denote as \textit{CE+ST} and \textit{CE+TT}, respectively.
Also, to evaluate the performance improvement brought by our proposed LG-Net model, we selected several CNN-based KWS models trained with CE loss functions for direct comparison~\cite{tang2018deep,li2020small,mo2020neural}.

\begin{table}
  \centering
  \setlength\tabcolsep{1pt}
  \caption{Comparison of the proposed method and the baselines}
    \begin{tabular}{lcccc}
    \hline
    Model & \multicolumn{1}{c}{Loss} & Params & Acc.(v1) & Acc.(v2) \\
    \hline
    \multirow{3}[1]{*}{\makecell[c]{CnnTradFpool3 \cite{sainath2015convolutional} \\ (Sainath \textit{et al.} 2015) }} & CE & \multirow{3}[1]{*}{1.39M} & 93.53\% & 90.15\% \\
      & CE+ST &   & 93.72\% & 90.19\% \\
      & CE+TT &   & 94.35\% & 91.45\% \\
    \cdashline{0-4}[0.8pt/2pt]
    \multirow{3}[0]{*}{\makecell[c]{CnnOneFstride4 \cite{sainath2015convolutional} \\ (Sainath \textit{et al.} 2015) }} & CE & \multirow{3}[0]{*}{3.84M} & 91.50\% & 89.16\% \\
      & CE+ST &   & 91.68\% & 90.03\% \\
      & CE+TT &   & 92.21\% & 90.38\% \\
    \cdashline{0-4}[0.8pt/2pt]
    \multirow{3}[0]{*}{\makecell[l]{TDNNSWSA \cite{bai2019time} \\ (Bai \textit{et al.} 2019)}} & CE & \multirow{3}[0]{*}{20K} & 94.19\% & 93.11\% \\
      & CE+ST &   & 95.32\% & 93.77\% \\
      & CE+TT &   & 95.47\% & 94.15\% \\
    \cdashline{0-4}[0.8pt/2pt]
    \multirow{3}[0]{*}{\makecell[l]{TC-ResNet8 \cite{choi2019temporal} \\ (Choi \textit{et al.} 2019) }} & CE & \multirow{3}[0]{*}{72K} & 95.71\% & 95.22\% \\
      & CE+ST &   & 96.12\% & 95.56\% \\
      & CE+TT &   & 96.58\% & 95.59\% \\
    \cdashline{0-4}[0.8pt/2pt]
    \multirow{3}[1]{*}{\makecell[l]{TC-ResNet14-1.5 \cite{choi2019temporal} \\ (Choi \textit{et al.} 2019) }} & CE & \multirow{3}[1]{*}{313K} & 96.78\% & 95.96\% \\
      & CE+ST &   & 96.81\% & 96.20\% \\
      & CE+TT &   & 96.89\% & \textbf{96.27}\% \\
    \cdashline{0-4}[0.8pt/2pt]
    \makecell[l]{Res15* \cite{tang2018deep} \\ (Tang \textit{et al.} 2018)} & CE &238K & 95.80\% & - \\
    \cdashline{0-4}[0.8pt/2pt]
    \makecell[l]{TENet12* \cite{li2020small} \\ (Li \textit{et al.} 2020)} & CE & 100K & 96.84\% & - \\
    \cdashline{0-4}[0.8pt/2pt]
    \makecell[l]{NAS2* \cite{mo2020neural} \\ (Mo \textit{et al.} 2020)} & CE & 886K & \textbf{97.22}\% & - \\
    \hline
    \multirow{3}[1]{*}{LG-Net3 (ours)} & CE & \multirow{3}[1]{*}{74K} & 95.71\% & 95.35\% \\
      & CE+ST &   & 96.46\% & 95.39\% \\
      & CE+TT &   & 96.82\% & 95.77\% \\
    \cdashline{0-4}[0.8pt/2pt]
    \multirow{3}[1]{*}{LG-Net6 (ours)} & CE & \multirow{3}[1]{*}{313K} & 97.03\% & 96.45\% \\
      & CE+ST &   & 97.13\% & 96.57\% \\
      & CE+TT &   & \textbf{97.67\%} & \textbf{96.79\%} \\
    \hline
    \multicolumn{5}{p{230pt}}{
    * represents the direct result of the corresponding paper.}
    \end{tabular}%
  \label{tab:result}%
\end{table}%

\subsection{Experimental Results and Analysis}
\begin{figure}
\centerline{\includegraphics[width=\columnwidth]{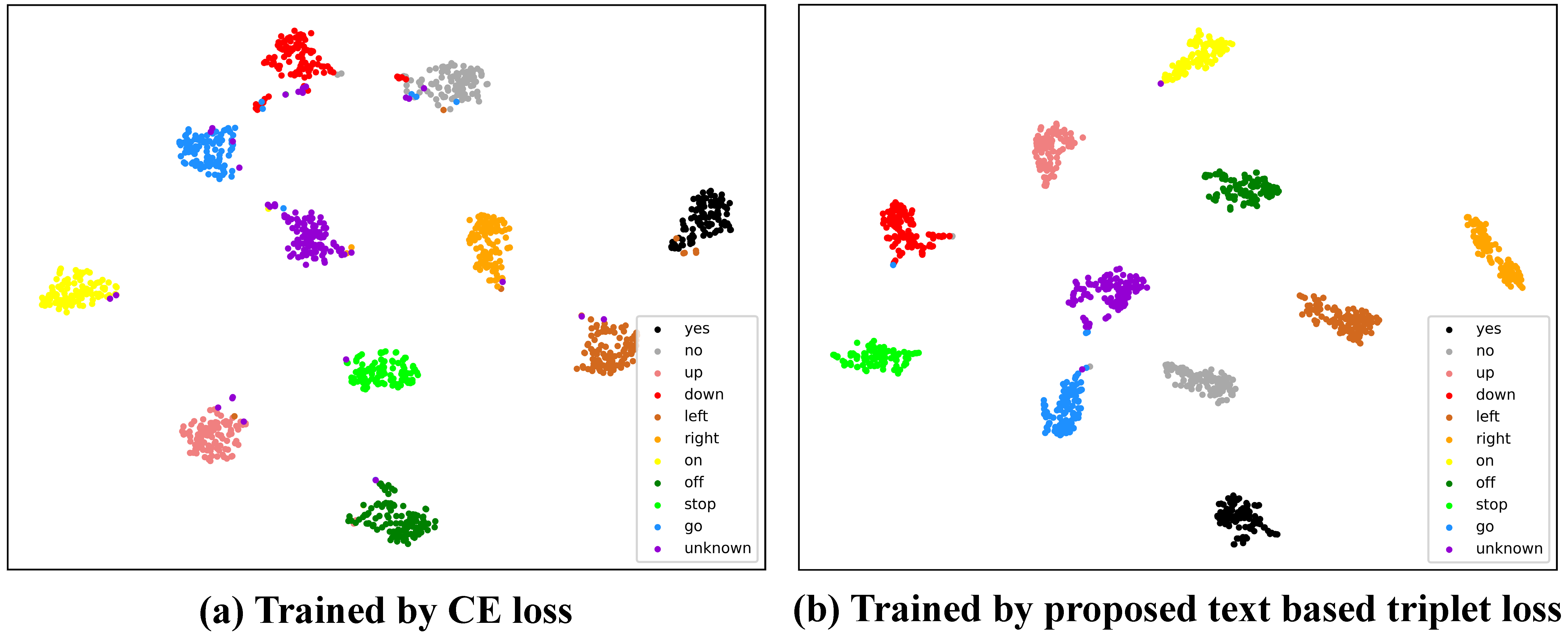}}
\caption{t-SNE visualization of speech embeddings extracted by LG-Net6.}
\label{cluster}
\end{figure}

\textbf{Impact of text anchor based metric learning method.} As Table \ref{tab:result} shows, for each model, training with \textit{CE+TT} loss shows consistent improvements over training with \textit{CE+ST} loss, since that BERT provides informative and robust text anchors, which alleviates the effects of the acoustic environment and speakers. Furthermore, We find that training with \textit{CE+ST} loss yields a performance gain compared to CE loss because triplet loss increases the distinguishability of features.
From Table \ref{tab:FRR}, we can see that LG-Net6 achieves a lower FRR at 0.5\% FAR. Note that there is no increase in memory footprint and computational resource at inference compared with the corresponding baseline.
As Figure \ref{cluster} shows, we randomly sample speech embeddings and use the t-SNE algorithm [23] to visualize the embeddings. We can see that keywords with similar pronunciation are difﬁcult to distinguish (e.g. ``go'' and ``no''), but are easily distinguished in the feature space after using our proposed text anchor based metric learning method.

\begin{table}
  \centering
  \caption{Comparison of FRR (false reject rate) of TC-ResNet14-1.5 and LG-Net6, FRR (false alarm rate) is at 0.5\%}
    \begin{tabular}{lccc}
    \hline
    Model & \multicolumn{1}{c}{Params} & \multicolumn{1}{c}{Loss} & \multicolumn{1}{c}{FRR} \\
    \hline
    \multirow{3}[2]{*}{TC-ResNet14-1.5} & \multirow{3}[2]{*}{313K} & CE & 5.91\%  \\
      & & CE+ST & 5.36\% \\
      & & CE+TT & 5.29\% \\
    \hline
    \multirow{3}[2]{*}{LG-Net6 (ours)} &  \multirow{3}[2]{*}{313K} & CE & 4.69\% \\
      & & CE+ST & 4.31\% \\
      & & CE+TT & \textbf{3.56}\% \\
    \hline
    \end{tabular}%
  \label{tab:FRR}%
\end{table}%

\textbf{Impact of other text vectors.} We make a step forward to study how other text vectors influence the performance of our proposed method. Comparison experiments among the different layers of BERT and Glove \cite{pennington2014glove} are performed. Referring to Table \ref{tab:addlabel}, both models, TC-ResNet14-1.5 \cite{choi2019temporal} and LG-Net6, achieve the best performances by employing the output of the first layer from BERT as the text anchor. The results indicate that the lower layer of BERT can mostly capture word-level information, which is consistent with the statement in \cite{jawahar2019does}.

\begin{table}
  \centering
  \caption{Comparison of other text embeddings on GSCDv2}
    \begin{tabular}{lcc}
    \hline
    Model & Text Embedding & Acc. \\
    \hline
    \multirow{4}[2]{*}{\makecell[l]{TC-ResNet14-1.5 \cite{choi2019temporal} \\ (Choi \textit{et al.} 2019) }} & glove.840B.300d \cite{pennington2014glove} & 96.21\% \\
      & BERT-Layer 1 \cite{devlin2018bert} & \textbf{96.27}\% \\
      & BERT-Layer 2 \cite{devlin2018bert} & 96.23\% \\
      & BERT-Layer 12 \cite{devlin2018bert} & 96.22\% \\
    \hline
    \multirow{4}[2]{*}{LG-Net6 (ours)} & glove.840B.300d \cite{pennington2014glove}& 96.47\% \\
      & BERT-Layer 1 \cite{devlin2018bert} & \textbf{96.79}\% \\
      & BERT-Layer 2 \cite{devlin2018bert} & 96.52\% \\
      & BERT-Layer 12 \cite{devlin2018bert} & 96.50\% \\
    \hline
    \end{tabular}%
  \label{tab:addlabel}%
\end{table}%

\textbf{Impact of the local-global block.} As shown in Table 2, compared to models trained using CE loss, LG-Nets achieve a balance between model size and accuracy. Compared to TC-ResNet14-1.5 \cite{choi2019temporal}, our LG-Net6 achieves better accuracy of 97.03\% and 96.45\% on GSCDv1 and GSCDv2, respectively. Compared to NAS2 [9], our proposed LG-Net6 achieves comparable results using only 35\% (313K/886K) of the number of parameters. The small-footprint model LG-Net3 obtains 95.71\% KWS accuracy in GSCDv1 and 95.35\% KWS accuracy in GSCDv2. Visualization results (Figure \ref{visual}b and Figure \ref{visual}c) show that the self-attention layer helps the network capture long-term temporal information while preserving the short-term temporal information captured by the 1D-CNN.

\section{Conclusion}
\label{sec:conclusion}
In this paper, we propose text anchor based metric learning method and design a neural network LG-Net for small-footprint keyword spotting (KWS).
The experimental results show that the proposed text anchor based metric learning method shows consistent improvements over the speech anchor based method. Moreover, our LG-Net model achieves SOTA accuracy on the Google Speech Commands Dataset version~1 and version~2.

\section{Acknowledgements}
This paper was partially supported by Shenzhen Science \& Technology Fundamental Research Programs (No: JSGG20191129105421211 \& GXWD20201231165807007-20200814115301001)

\bibliographystyle{IEEEtran}

\bibliography{mybib}


\end{document}